# Hierarchical Bayesian myocardial perfusion quantification

Cian M. Scannell, MRes[1,2]; Amedeo Chiribiri, MD, PhD[1]; Adriana D.M. Villa, MD, PhD[1]; Marcel Breeuwer, PhD[3,4]; Jack Lee, DPhil[1];

[1]School of Biomedical Engineering and Imaging Sciences, King's College London, United Kingdom

[2]The Alan Turing Institute London, United Kingdom

[3]Philips Healthcare, Best, The Netherlands

[4]Department of Biomedical Engineering, Medical Image Analysis group, Eindhoven University of Technology, Eindhoven, The Netherlands

cian.scannell@kcl.ac.uk

amedeo.chiribiri@kcl.ac.uk

adriana.villa@kcl.ac.uk

marcel.breeuwer@philips.com

jack.lee@kcl.ac.uk

Corresponding author:

Cian M. Scannell, School of Biomedical Engineering and Imaging Sciences, King's College London 4th Floor Lambeth Wing, St Thomas' Hospital, Westminster Bridge Road, London SW1 7EH 9, United Kingdom

# Abstract

*Purpose:* Tracer-kinetic models can be used for the quantitative assessment of contrast-enhanced MRI data. However, the model-fitting can produce unreliable results due to the limited data acquired and the high noise levels. Such problems are especially prevalent in myocardial perfusion MRI leading to the compromise of constrained numerical deconvolutions and segmental signal averaging being commonly used as alternatives to the more complex tracer-kinetic models.

*Methods*: In this work, the use of hierarchical Bayesian inference for the parameter estimation is explored. It is shown that with Bayesian inference it is possible to reliably fit the two-compartment exchange model to perfusion data. The use of prior knowledge on the ranges of kinetic parameters and the fact that neighbouring voxels are likely to have similar kinetic properties combined with a Markov chain Monte Carlo based fitting procedure significantly improves the reliability of the perfusion estimates with compared to the traditional least-squares approach. The method is assessed using both simulated and patient data.

*Results*: The average (standard deviation) normalised mean square error for the distinct noise realisations of a simulation phantom falls from 0.32 (0.55) with the least-squares fitting to 0.13 (0.2) using Bayesian inference. The assessment of the presence of coronary artery disease based purely on the quantitative MBF maps obtained using Bayesian inference matches the visual assessment in all 24 slices. When using the maps obtained by the least-squares fitting, a corresponding assessment is only achieved in 16/24 slices.

*Conclusion*: Bayesian inference allows a reliable, fully automated and user-independent assessment of myocardial perfusion on a voxel-wise level using the two-compartment exchange model.

# 1. Introduction

Dynamic contrast-enhanced magnetic resonance imaging (DCE-MRI) can be used for the non-invasive assessment of myocardial perfusion (Chiribiri et al., 2009; Jaarsma et al., 2012; Nagel et al., 2003). According to recent clinical guidelines, it is indicated for the assessment of patients at risk of coronary artery disease (CAD) (Montalescot et al., 2013; Windecker et al., 2014) and has been extensively validated against the reference standard, fractional flow reserve (Li et al., 2014). Currently, the clinical evaluation of such image series is performed visually. The spatial and temporal distribution of contrast agent in the myocardium can identify myocardial ischaemia and provide insight into the presence and severity of stenosis. Some of the main limitations of this visual assessment are the difficulty of interpreting the images (Villa et al., 2018) and the underestimation of the ischaemic burden in patients with multivessel CAD (Patel et al., 2010). This has led to myocardial perfusion examinations only being routinely performed in highly experienced centres. Quantitative perfusion analysis has been proposed as a more reproducible and user-independent alternative of visual assessment and has been shown to have a good diagnostic accuracy and prognostic value (Hsu et al., 2018; Knott et al., 2019; Sammut et al., 2017).

The quantification of myocardial perfusion from DCE-MRI data is achieved by applying tracer-kinetic models to track the passage of the contrast agent from the left ventricle (LV) to the myocardium and infer the kinetic model parameters, such as myocardial blood flow (MBF) (Broadbent et al., 2013; Hsu et al., 2018; M Jerosch-Herold et al., 1998; Kellman et al., 2017; Wilke et al., 1997). However, there has been questions raised on the reliability of the quantitative parameters values that are estimated from DCE-MRI data due to the complexity of the models relative to the observed data (Buckley, 2002). This had led to the use of simplified models, such as the Fermi function (Jerosch-Herold et al., 1998; Wilke et al., 1997) or concentration curves that have been averaged over a region of the myocardium in order to boost the signal-to-noise ratio (SNR). A recent editorial by Axel (Axel, 2018) calls for improved quantitative methods, in particular more robust quantitative values.

Larsson et al. (Larsson et al., 1996) showed that they could not reliably fit the five parameters needed for the tracer-kinetic modelling to the observed data, using least-squares fitting. As shown in the comparative study of Schwab et al. (Schwab et al., 2015), they were able to achieve reliable quantification with relatively simpler models, such as a Fermi-constrained deconvolution, but not with the multi-compartment exchange models. Broadbent et al. (Broadbent et al., 2013) report failed fitting in 10% of cases (despite using concentration curves that have been averaged over a segment of the myocardium) and Likhite et al. (Likhite et al., 2017) also reported failed fittings to simulated data which is simplistic with respect to the patient data.

Some of the reasons behind the reported difficulties in the model fitting include that such parameter estimation, or non-linear regression, problems are known to get stuck in local optima (Dikaios et al., 2017). As a result, even though the model-based concentration curves may well match the observed data, the reported parameters may be far from the true values. It has further been shown that the model parameters are correlated (Romain et al., 2017) and thus there are multiple distinct combinations of parameters that give outputs that are indistinguishable at the observed noise level. Also, as is typical with non-linear optimisations, the parameter estimates are highly sensitive to the initial conditions of the optimisation and the specific noise present in the data. A further limitation is that this non-linear least-squares fitting does not explicitly deal with the uncertainty in the estimated kinetic parameters.

In conclusion, there is need for an improved methodology to allow robust and reproducible estimation of the kinetic model parameters including, but not limited to, MBF. In this work, we develop and evaluate a framework to robustly infer the kinetic model parameters from the observed imaging data based on hierarchical Bayesian probabilistic modelling. This approach is validated using simulation phantoms where gold standard kinetic parameters are known and subsequently further testing on clinical data is reported.

## 2. Background

*2.1 Tracer-kinetic models*

The tracer-kinetic models as presented in the literature (Ingrisch and Sourbron, 2013; Sourbron and Buckley, 2013) model the perfusion unit (a single voxel or segment) as a system with two interacting compartments - the plasma and the interstitium. These models give a pair of coupled differential equations which describe the evolution of the contrast agent as a non-linear function of physiological parameters, such as MBF. In this work, the tracer-kinetic model analysis was performed by fitting a two-compartment exchange model (2CXM) (Jerosch-herold, 2010; Sourbron and Buckley, 2013) to the observed concentration curves

$$v_p \frac{dC_p(t)}{dt} = \frac{F_b}{1-Hct}\left(C_{AIF}(t) - C_p(t)\right) + PS\left(C_e(t) - C_p(t)\right) \quad (1)$$

$$v_e \frac{dC_e(t)}{dt} = PS\left(C_p(t) - C_e(t)\right). \quad (2)$$

In (1) and (2), $C_p(t)$ and $C_e(t)$ are the concentration of contrast agent in the plasma and interstitial space at time $t$, respectively (in units of M). $C_{AIF}(t)$, the arterial input function (AIF), is the assumed input to the system that is being modelled (also in units of M). In myocardial perfusion quantification this is sampled from the LV. $F_b$ is the MBF (mL/min/mL), $v_p$ is the fractional plasma volume (dimensionless), $v_e$ is the fractional interstitial volume (dimensionless) and $PS$ is the permeability-surface area product (mL/min/mL). $Hct$ is the haematocrit value (dimensionless).

This model has the benefit over other simpler models in that it resolves directly for MBF. The simpler models, such as those presented by Tofts et al. (Tofts and Kermode, 1991), only allow the estimation of the $K^{trans}$ parameter which can be influenced by either MBF or the extraction fraction. The Fermi function (Jerosch-herold et al., 1998; Wilke et al., 1997) does resolve for MBF but not other kinetic parameters and the model is not physiologically motivated.

The solution to this system is then given as:

$$C_\Theta(t) = R_F(t, \Theta) * C_{AIF}(t - \tau_0) \quad (3)$$

with the analytic form of the residue function $R_F$ presented in the appendix. $\boldsymbol{\Theta} = (F_p, v_p, v_e, PS)^T$ and $\tau_0$ is the time delay term which accounts for the fact that the contrast agent does not move instantaneously from the left ventricle to the myocardial tissue. This is an unknown parameter that must also be estimated. The concentration that is observed in the MRI experiment is the contribution from both compartments and is given as: $C(t) = v_p \cdot C_p(t) + v_e \cdot C_e(t)$.

*2.2 Non-linear regression*

The standard technique to estimate the model parameters uses a least-squares method. Given the observed contrast agent concentrations $\boldsymbol{y} = (y(t_0), y(t_1), \ldots, y(t_{N-1}))^T$, it is assumed that: $y(t_j) = C_{\boldsymbol{\Theta}}(t_j) + \epsilon_j$ where $\epsilon_j$ are the error terms and comprise of both noise and other sources of error, such as motion. The estimation of the parameters is then to find the $\boldsymbol{\Theta}$ which minimises the sum of squared errors cost function $\chi^2$:

$$\widehat{\boldsymbol{\Theta}} = argmin_{\boldsymbol{\Theta}} \chi^2(\boldsymbol{\Theta}) = argmin_{\boldsymbol{\Theta}} \frac{1}{N} \sum_{j=0}^{N-1} \left( C_{\boldsymbol{\Theta}}(t_j) - y(t_j) \right)^2 \qquad (4)$$

Under the assumption that the error terms come from independent and identically distributed Gaussian distribution this is equivalent to the maximum likelihood estimate as it maximises the likelihood function $p(\boldsymbol{y}|\boldsymbol{\Theta})$.

As previously discussed, this technique can break down in the case where the cost function has multiple local minima. That is, there are multiple values of $\boldsymbol{\Theta}$ that produce similar model output. If this is the case, the values of the parameters estimated may depend strongly on the initial conditions of the optimisation and be far from the true values. Furthermore, *in vivo,* the time delay parameter $\tau_0$ can introduce further local minima.

An example of such a cost function is shown in Fig. 1. It is seen that when noise is added, two local optima emerge, neither of which correspond to the true parameter values. The optimisation will converge to one of these depending on its initial conditions. In other cases it is possible for the cost functions to possess long flat valleys where optimisation may stop due to the update being less than the required tolerance leading to unreliable parameter estimates. This cost function was constructed using simulated data and can possess further local optima due to the complex errors and physiology seen in patient data. Additionally, as shown by Sommer and Schmid (Sommer and Schmid, 2014), the analytic form of the residue function $R_F$, which is the sum of two exponentially decaying components, can lead to an identifiability problem when the two exponents are too similar or when the contribution of one compartment vanishes which can further reduce the reliability of the parameter estimates.

**Figure 1.**

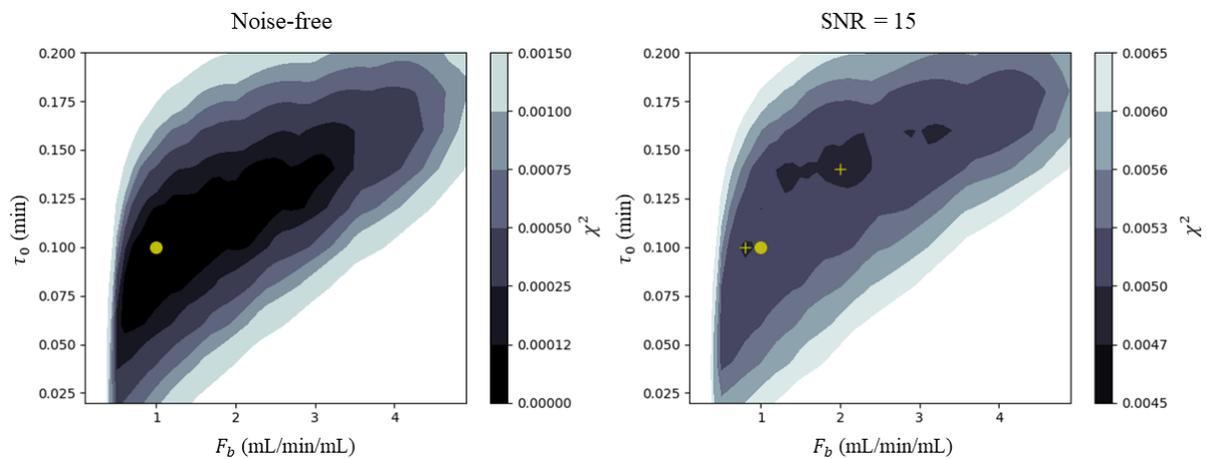

*An example of a cost function created using simulated data, on the left in the noise-free case and on the right with an application realistic SNR of 15. This image is constructed by taking the minimum intensity projection over the three parameters $v_p, v_e$ and $PS$ to allow the visualisation of the 5-dimensional surface as a function of $F_b$ and $\tau_0$. The true parameter values used in the forward simulation are $F_b = 1.0, v_p = 0.08, v_e = 0.16, PS = 0.4, \tau_0 = 0.1$. In the noise-free (left) case the cost function has a global minimum which corresponds to the true parameter values (yellow dot). In the presence of noise (right) it is seen that there are two local optima, neither of which correspond to the true parameter values. The optimisation will converge to one of these depending on its initial conditions. The yellow circle is the position of the true minimum of the cost function and the two cross symbols are the positions of the two local minima.*

Fig. 2 shows two myocardial tissue curves that have been simulated using the same parameters, with the only difference being the realisation of the Rician noise that is added. This could be interpreted as being two curves from neighbouring voxels with the same underlying physiology. In this example it is seen how the traditional non-linear regression algorithms can yield vastly different fits, with the two fitted MBF values ($F_b$) being different by a factor of two despite being simulated with the same kinetic parameters.

**Figure 2.**

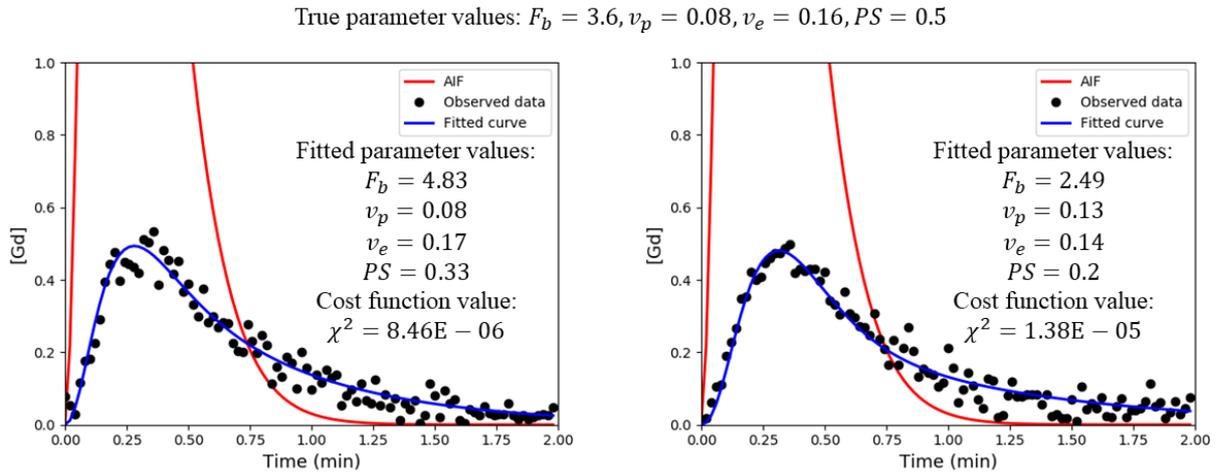

*For the same arterial input function AIF (red curves) and the same ground-truth parameters ($F_b = 3.6, v_p = 0.08, v_e = 0.16, PS = 0.5$) the two blue curves are simulated. Their only difference being the Rician noise realisation. A comparison of the two fits shows a difference of a factor of two in the computed MBF. This could vastly change the patients' diagnosis.*

*2.3 Bayesian parameter estimation*

The aforementioned maximum-likelihood approach assumes that there is one true value of the parameter and computes a point estimate of this. Conversely, Bayesian estimation treats the parameters as random variables and approximates their posterior distribution. This hence allows computation of the expected value of the parameter. The variance of the distribution also allows an expression of confidence in the value of the parameter estimate.

It is assumed that the observed data $y$ at each time point $t_j$, $j = 0, \dots, N-1$ comes from the model with some Gaussian error with variance $\sigma_j^2$ such that: $y(t_j) \sim \mathcal{N}(C_\Theta(t_j), \sigma_j^2)$. To then examine to model parameters given observed data, the posterior distribution $p(\Theta|y)$ is required. The posterior distribution can be obtained through the application of Bayes' theorem

$$p(\mathbf{\Theta}|\mathbf{y}) = \frac{p(\mathbf{y}|\mathbf{\Theta}) \cdot p(\mathbf{\Theta})}{\int_{\mathbf{\Theta}} p(\mathbf{y}|\mathbf{\Theta}) \cdot p(\mathbf{\Theta}) \, d\mathbf{\Theta}} \propto p(\mathbf{y}|\mathbf{\Theta}) \cdot p(\mathbf{\Theta}). \tag{5}$$

The term $p(\mathbf{\Theta})$ is the prior probability of the parameters. It is common to assume that the parameters $\theta_i$ are independent and so $p(\mathbf{\Theta})$ is the product of the prior distributions over the individual parameters, $\theta_i$.

In general, the posterior distribution is not analytically tractable and must be approximated using Markov chain Monte Carlo (MCMC) sampling. Through (5), it is possible to compute samples which are proportional to the posterior distribution and MCMC sampling utilises these samples to construct a Markov chain with a stationary distribution equal to the posterior distribution. It does this using a proposal distribution which is used to propose, randomly, how to move in parameter space and an acceptance rule which is used to decide whether to accept the proposed move or not, based on the information from the likelihood and the prior information. An introduction to the use of Bayesian modelling for non-linear regression problems is given in the book of Seber and Wild (Seber and Wild, 1989).

*2.3.1 Hierarchical Bayesian modelling*

Quantitative myocardial perfusion MRI is a natural application for the use of hierarchical Bayesian modelling. In this modelling approach, the prior probability distribution is not governed by fixed hyperparameters but rather hyperparameters $\boldsymbol{\alpha}$ which are described by a further probability distribution, i.e a hyperprior $p(\boldsymbol{\alpha})$. Hence, $\mathbf{\Theta}$ in (5) is now dependent on these hyperparameters and (5) becomes:

$$p(\mathbf{\Theta}, \boldsymbol{\alpha}|\mathbf{y}) \propto p(\mathbf{y}|\mathbf{\Theta}, \boldsymbol{\alpha}) \cdot p(\mathbf{\Theta}|\boldsymbol{\alpha}) \cdot p(\boldsymbol{\alpha}) \tag{6}$$

This approach is useful when the data is structured into distinct but not entirely unrelated groups. This is referred to as partial pooling, as opposed to complete pooling (use of one fixed prior distribution) or no pooling (use of different priors for each group). For example, in a stress perfusion MRI, these distinct groups would be healthy and diseased tissues. If the same prior knowledge was used for both groups, then it would lead to an averaging effect over these regions. Hierarchical modelling is thus an attractive compromise between treating the groups equivalently and having completely independent models.

*2.3.2 Generalised Gaussian Markov random field prior*

In addition to prior distributions on the kinetic parameters, it is possible to incorporate spatial prior knowledge. This enforces smoothness in the spatial domain and it motivated by the idea that neighbouring voxels should exhibit similar kinetic properties. In particular, in this application, a generalised Gaussian Markov random field prior is suitable. Mathematically, this is equivalent to putting prior distributions on the differences between parameters in neighbouring voxels that have zero mean:

$$\mathbf{p}(\mathbf{\Theta}_i|\mathbf{\Theta}_j, \nu_{i,j}) \propto \exp(-\frac{\nu_{i,j}}{2} \|\mathbf{W}(\mathbf{\Theta}_i - \mathbf{\Theta}_j)\|_p^p) \text{, if } j \sim i \qquad (7)$$

where $j \sim i$ if $j$ and $i$ are neighbouring voxels and $1 \leq p \leq 2$. The use of $p = 1$ corresponds to the Laplace distribution and is known to have edge-preserving properties (Bardsley, 2012).

## 3. Methods

*3.1 Simulation experiments*

The proposed method was first tested using simulated image series as the parameter estimates can be compared to ground-truth values. The 6 by 6 voxel image series was created using ground truth tracer-kinetic parameter maps with values as expected in the myocardium under three different realistic conditions mimicking a healthy patient at rest, a healthy patient at stress and a patient with stress-inducible ischaemia (Broadbent et al., 2013). The parameter maps were used to forward simulate the model with a gamma-variate function used to generate a realistic AIF. The kinetic parameter values used in the simulation were $F_b = 3.5, v_p = 0.08, v_e = 0.16, PS = 1.0$ in healthy voxels at stress and $F_b = 1.0, v_p = 0.08, v_e = 0.16, PS = 1.0$ in healthy voxels at rest. The simulation phantom mimicking a patient with stress-inducible ischaemia was created using $F_b = 3.5, v_p = 0.08, v_e = 0.16, PS = 1.0$ with two disconnected regions with reduced MBF ($F_b = 1.0$) added to mimic regions of stress-inducible myocardial ischaemia. Rician noise was added to the image, with the level chosen to achieve a realistic SNR of 15 (Broadbent et al., 2016; Cheng et al., 2007). The SNR here was defined to be the ratio of the standard deviation of the noise realisation to the maximum of the tissue curves. The data curves were simulated with a realistic temporal resolution of $\Delta t = 0.012$ min, corresponding to a heart rate of roughly 83 beats per minute at stress and $\Delta t = 0.017$ min at rest, for a total time $T = 3$ min. The proposed parameter estimation method is compared in a Monte-Carlo study for $n = 20$ distinct noise realisations, for each simulation phantom, with a traditional, gradient-based optimisation scheme in order to assess the accuracy and the reproducibility of the parameter estimates. The normalised mean square error (NMSE) between the true and estimated kinetic parameters is reported and a Mann-Whitney U test is used to compare the distribution of the NMSE values from the Monte-Carlo study. A further assessment is conducted (also using the NMSE values) to compare the proposed hierarchical model to an equivalent non-hierarchical approach.

*3.2 In vivo experiments*

The technique was tested in eight patients suspected of having CAD referred for stress perfusion cardiac MRI at King's College London. Image acquisition was performed on a 3.0T scanner (Philips Achieva-TX, Philips Medical Systems) using standard acquisition protocols (Kramer et al., 2013). The typical acquisition parameters, TR/TE/flip angle/saturation prepulse delay were 2.5 ms/1.25 ms/15° /100 ms with a typical spatial resolution of 1.34 x 1.34 x 10 mm. The dynamic image series were acquired during first-pass injection of 0.075 mmol/kg Gadobutrol (Gadovist, Schering, Germany) at 4 ml/s followed by a 20 ml saline flush. A dual bolus contrast agent scheme was used to correct for signal saturation of the AIF, as previously described (Ishida et al., 2011). Images were acquired under adenosine-induced stress. The images were motion corrected using a previously validated scheme (Scannell et al., 2019).

As aforementioned, a dual-bolus acquisition is performed in order to mitigate the difficulties caused by the non-linear relationship between the concentration of contrast agent and the MRI signal intensities. It is hence assumed there is a linear relationship between the concentration of contrast agent and the signal intensity. The concentration of gadolinium (C(t)) was approximated from the signal intensities (S(t)) using an application specific version of the relative signal enhancement (Biglands et al., 2015; Ingrisch and Sourbron, 2013):

$$C(t) = \frac{1}{r_1 \cdot T_{1b}} \left( \frac{S(t) - S(0)}{S_{LV}(0)} \right) \qquad (8)$$

with the $T_{1b}$ of blood taken as 1736 ms and $r_1$ the contrast agent as 4.5 s$^{-1}$ mmol/L$^{-1}$ (Broadbent et al., 2016). S(0) is the average of the first five acquired images before the injection of contrast agent. Similarly, $S_{LV}(0)$ is the pre-contrast signal in the left ventricular blood pool.

In the case of this patient data, there are no ground-truth parameter values to compare to. Therefore, the purpose of this study is to test whether the kinetic parameter values that are estimated can identify perfusion defects that match the visual assessment found in the clinical reports. This provides insight into the reliability of the fittings and the ability of the proposed method to deal with the more complex curves and error terms that are present in patient data. The number of failed fittings and outlier kinetic parameter estimates are further compared between implementations. All scans were reported blindly by experienced operators with level III CMR accreditation according to the guidelines of the Society for Cardiovascular Magnetic Resonance (SCMR). As per the expert assessment, four patients were classified as being positive for ischaemia and four were classified as not having obstructive coronary artery disease. The myocardium was contoured using the cvi$^{42}$ software (Circle Cardiovascular Imaging Inc., Calgary, Alberta, Canada) by an experienced operator with level III CMR accreditation (SCMR) and the segmentations were exported using the open-source code of Bai et al. (Bai et al., 2017)

*3.3 Non-linear regression implementation*

All steps are implemented in Python, using the SciPy module for the optimisation (Jones et al., n.d.). The nonlinear regression approach uses the L-BFGS (Zhu et al., 1997) nonlinear optimisation scheme with box constraints. Each parameter was constrained to be within conservative physiological limits and to conform with what has been previously found with tracer-kinetic models (Broadbent et al., 2013). The parameters are constrained such that: $0.001 \leq F_b \leq 6.0$, $0.001 \leq v_p \leq 0.3$, $0.001 \leq v_e \leq 0.4$ and $0.001 \leq PS \leq 4.0$. This fitting is repeated 100 times with different initial conditions randomly chosen from a uniform distribution on each of these ranges (Romain et al., 2017). The reported parameter estimates are then the successful fit which has achieved the lowest cost function value. This is done to reduce the effect of the choice of the initial conditions on the parameter estimate and to minimise the risk of converging to local optima. A fit is defined as successful if it achieves a tolerance of less than $10^{-8}$ within 1000 iterations and none of the resulting parameters achieve their upper or lower bounds. The AIF ($C_{AIF}(t)$) is extracted using independent component analysis (Jacobs et al., 2016) and the bolus arrival time is estimated using the method of Cheong et al (Cheong et al., 2003).

*3.4 Bayesian parameter estimation implementation*

The Bayesian parameter estimate was implemented using an in-house software developed in Python. The posterior distribution for the parameters in voxel $i$ is given through the application of Bayes' theorem as:

$$p(\mathbf{\Theta^i}, \mathbf{\alpha^i}|\mathbf{y^i}) \propto p(\mathbf{y^i}|\mathbf{\Theta^i}, \mathbf{\alpha^i}) \cdot p(\mathbf{\Theta^i}|\mathbf{\alpha^i}) \cdot p(\mathbf{\alpha^i}) \qquad (9)$$

It is assumed that the observed data in voxel $i$ at time $t_j$ is Gaussian distributed, i.e that $y^i(t_j) \sim \mathcal{N}(C_{\Theta^i}(t_j), \sigma_i^2)$. This gives rise to the likelihood function:

$$p(\mathbf{y^i}|\Theta^i, \alpha^i) = (2\pi\sigma_i^2)^{-\frac{N}{2}} \exp\left(-\frac{1}{2\sigma_i^2} \sum_{j=0}^{N-1} \left(y^i(t_j) - C_{\Theta^i}(t_j)\right)^2\right) \quad (10)$$

In this work, $F_b$ (mL/min/mL) is selected to be Gaussian distributed with mean $\alpha_p$ and a fixed variance 0.2. $PS$ (mL/min/mL) is Gaussian distributed with mean $\alpha_S$ and variance 0.1. $\alpha_p$ is taken to be uniformly distributed on [0,7] and $\alpha_S$ is taken to be uniformly distributed on [0,5]. $v_p$ (%/100) is assumed to be uniformly distributed on [0,0.4] and $v_e$ (%/100) is assumed to be uniformly distributed on [0,0.5]. These were chosen to be in line with previously reported literature values (Broadbent et al., 2013) and physiological intuition. The priors are chosen to be weakly informative in that they encompass a much larger range of values than the values that have been found previously in the literature. Rather than expressing a confidence about the parameters close to a certain value, it restricts the parameter estimates to these ranges. The prior distribution on the observed error (in M) for voxel $i$ ($\sigma_i^2$) is taken to be a flat Inverse-Gamma distribution, with shape parameter $c = 0.001$ and scale parameter $d = 0.001$, as is conventional.

A Laplace prior with location 0 and scale 0.2 is chosen on the absolute value of the distance between the kinetic parameter estimates of neighbouring voxels. The Laplace distribution is chosen due to its edge preserving properties. This gives rise to the prior distribution:

$$p(\Theta^i|\alpha^i) = p\left(F_p^i \middle| \alpha_{F_p}^i\right) \cdot p(v_p^i) \cdot p(v_e^i) \cdot p(PS^i|\alpha_{PS}^i) \cdot p(\sigma_i^2) \cdot p(\Theta^i|\Theta^{n(i)}, \alpha^i, v_{i,j})$$

$$\propto \exp\left(-\frac{1}{2 \cdot 0.1}\left(F_p^i - \alpha_{F_p}^i\right)^2\right) \times \mathbb{I}(v_p^i \in (0,0.3]) \times \mathbb{I}(v_e^i \in (0,0.4]) \times$$

$$\exp\left(-\frac{1}{2 \cdot 0.1}(PS^i - \alpha_{PS}^i)^2\right) \times \left(\frac{1}{\sigma_i^2}\right)^{c-1} \exp\left(-\frac{d}{\sigma_i^2}\right)$$

$$\times \exp\left(-\frac{1}{0.2} \sum_{j \in n(i)} \sum_{k=1}^{4} (W_k \cdot |\Theta_k^i - \Theta_k^j|)\right) \quad (11)$$

𝕀(X) is the indicator function on the set X which takes the value 1 on X and 0 otherwise. $n(i)$ is the set of neighbouring voxels of voxel $i$. A voxel's neighbours are those voxels in its surrounding 4-neighbourhood, above, below, to the left and to the right of it. Due to the shape of the myocardium, it is possible that a voxels neighbours are not in the myocardial segmentation. In such a case, a voxel diagonally above or below is taken or failing that, the closest voxel that is in the myocardium.

The distance between the parameter estimates in neighbouring voxels $i$ and $j$ is computed using a weighted sum. The weights, $W_k$ are used to account for the different scales of the parameters, since otherwise, differences in the higher magnitude parameter values ($F_b$ and $PS$) would have a dominating effect compared to the lower magnitude parameter values ($v_p$ and $v_e$). The value of the weight for a given parameter $W_k$ on a given iteration of the MCMC sampling is the inverse of the previous sample of the parameter. For the non-hierarchical model it is taken that $F_b \sim \mathcal{N}(3.5, 0.2)$ and $PS \sim \mathcal{N}(1.0, 0.2)$.

The hyperprior distribution is given as:

$$p(\boldsymbol{\alpha^i}) = \mathbb{I}\left(\alpha_{F_p}^i \in [0.001, 7]\right) \times \mathbb{I}\left(\alpha_{PS}^i \in [0.001, 5]\right) \quad (12)$$

In this work, the Metropolis-Hasting algorithm with random walk proposals is used to sample from the posterior distribution. In short, the Metropolis-Hasting algorithm moves randomly through parameter space using a proposal distribution. The proposal distribution proposes a move in parameter space that is then accepted with a probability that is related to the change in posterior probability associated with the new sample. This leads to the construction of a Markov chain with a stationary distribution that approximates the posterior distribution.

In the work, the choice of the above distributions and their respective variances was made, empirically, to optimise the trade-off between thoroughly exploring parameter space and sticking to areas with high posterior probability. They were chosen in order to achieve a rate of acceptance of proposals close to 0.234 which has previously been shown to be optimal (Roberts et al., 1997). Markov chains of 4000 steps were constructed. In order to assess the stationary distribution of the chain, the initial 1000 steps were discarded, referred to as the burn-in phase. The number of steps was chosen to be far in excess of the number needed for convergence according to the $\hat{R}$ statistic (Gelman and Rubin, 1992). In order to create parameter maps, the median value of the posterior distribution are reported and to examine the uncertainty associated with such a parameter estimate, the coefficient of variation of the posterior distribution, the ratio of the standard deviation of the distribution to its mean value, is reported.

# 4. Results

*4.1 Simulations*

Table 1 shows the mean (standard deviation) NMSE between the estimated and true kinetic parameters values for both the hierarchical Bayesian and non-linear least squares implementations, with the results of the Mann-Whitney U-test. The NMSE is significantly lower for the Bayesian method compared to the non-linear least squares. Example parameter maps from both methods are compared to the true parameter maps in Fig. 3. A significantly higher NMSE was found using in the stress simulations with a perfusion defect using the non-hierarchical approach (0.24 (0.15), $p < 0.001$) with a comparison of the computed MBF parameter maps for an example noise realisation shown in Fig. 4.

**Table 1.**

| Parameter | Bayesian | Non-linear least squares | p-value Mann-Whitney U test |
|:---:|:---:|:---:|:---:|
| All | 0.13 (0.2) | 0.32 (0.55) | $p < 0.0001$ |
| $F_b$ | 0.05 (0.09) | 0.1 (0.09) | $p = 0.002$ |
| $v_p$ | 0.22 (0.27) | 0.35 (0.31) | $p = 0.02$ |
| $v_e$ | 0.12 (0.16) | 0.20 (0.17) | $p = 0.01$ |
| PS | 0.11 (0.21) | 0.63 (0.96) | $p < 0.0001$ |

*Table 1: NMSE and Mann-Whitney U test results for the hierarchical Bayesian and non-linear least squares kinetic parameter estimates.*

**Figure 3.**

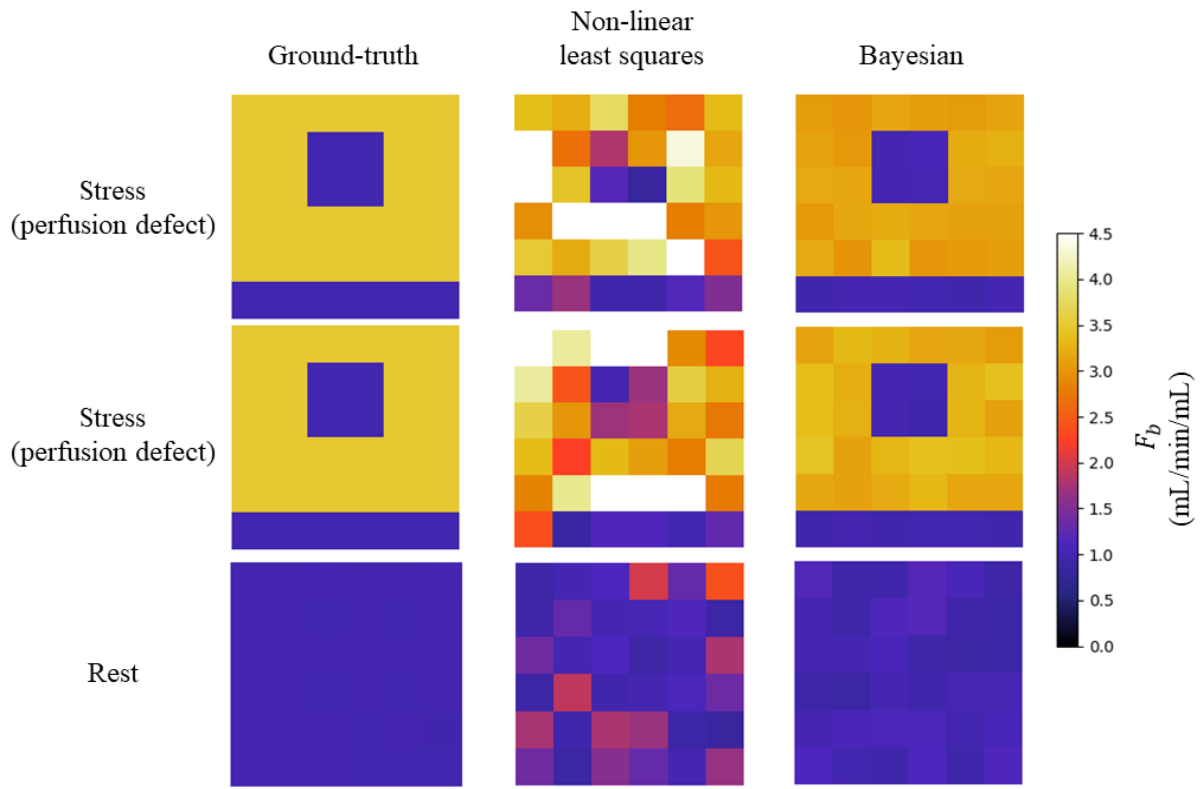

*The ground-truth $F_b$ parameter values (left) are compared to values that are estimated using non-linear least-squares fitting (middle), and the proposed Bayesian inference method (right) for two random noise realisations of the simulations mimicking stress-inducible ischaemia (top and middle) and one random noise realisation of the rest simulations (bottom). The Bayesian inference is significantly closer to the ground-truth with fewer outliers.*

**Figure 4.**

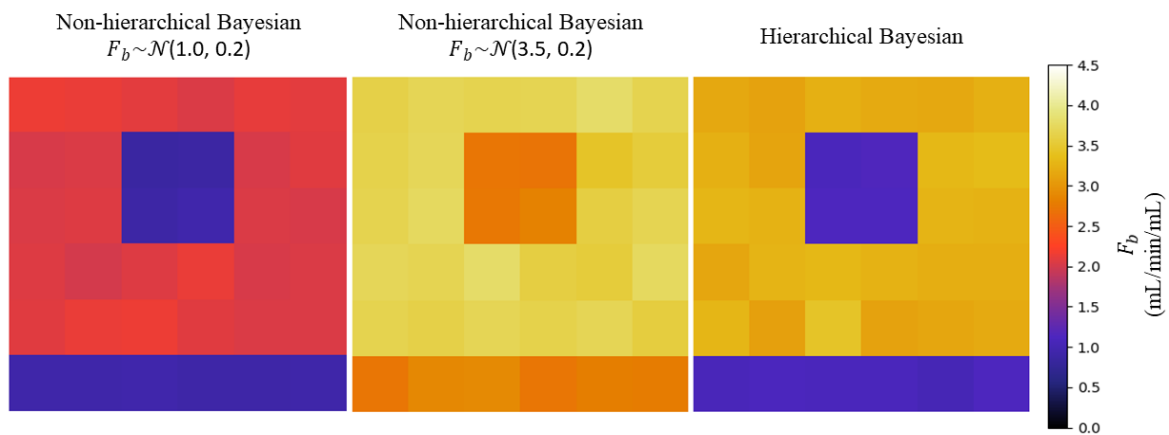

*The comparison between the hierarchical and non-hierarchical approach for an example noise realisation with the same ground-truth MBF maps at stress with a simulated perfusion defect as shown in Fig. 3. This shows the effect of using prior distributions with fixed means which influence the information from the data to drive the parameter estimate towards the prior value.*

*4.2 Patient data*

The median computed MBF value (25$^{th}$ percentile, 75$^{th}$ percentile) was 2.35 (1.9, 2.68) mL/min/mL under stress conditions using the proposed Bayesian inference scheme. The equivalent results were 2.37 (1.12, 3.01) mL/min/mL using the non-linear least squares fitting. However, with the least squares fitting approach there is a number of voxels for which the fitting fails completely, which are represented as holes in the parameter maps, as seen in Fig. 5. The proposed Bayesian inference techniques has zero voxels with estimates converging to upper or lower bounds or outside physiological ranges. The least-squares fitting fails for an average (standard deviation) of 12.9% (12.4%) of voxels per slice. Additionally, a MBF value of greater than 5 mL/min/mL (considered to be outliers) was found in 7.5% of voxels using least-squares fitting but never achieved with the Bayesian inference. The other kinetic parameter values are quoted in Table 2. The median (25$^{th}$ percentile, 75$^{th}$ percentile) coefficient of variation of the Bayesian posterior was 6.6% (3.3%, 11.7%). A maximum value of 87.7% was achieved with 0.8% of voxels having a parameter with a coefficient of variation greater than 50%.

The assessment of the presence of coronary artery disease based purely on the quantitative flow maps obtained using Bayesian inference matches the visual assessment in all 24 slices. When using the maps obtained by the least-squares fitting, a corresponding assessment is achieved in 16/24 slices. The computed flow parameter maps under stress conditions for an example patient with a perfusion defect are shown in Fig. 5. The identified areas of ischaemia are indicated with a blue arrow in the original MR images. An example of a slice where the least-squares fitting fails to correspond to the visual assessment is shown in Fig. 6. The visual assessment concluded that there is reduced uptake of contrast agent in both the inferior and infero-septal segments. This clearly corresponds with the Bayesian inference. The least-squares fitting is extremely noisy and the ischaemic area is under-estimated in the inferior segment and almost completely missed in the infero-septum. Fig. 7. shows all four kinetic parameters (left column) with the coefficient of variation of the MCMC sample of the parameter posterior distribution (right column).

**Table 2.**

| *Parameter* | **Bayesian** | **Non-linear least squares** |
|---|---|---|
| $F_b$ (mL/min/mL) | 2.35 (1.9, 2.68) | 2.37 (1.12, 3.01) |
| $v_p$ (%/100) | 0.09 (0.05, 0.13) | 0.07 (0.03, 0.13) |
| $v_e$ (%/100) | 0.21 (0.13.0.31) | 0.19 (0.12, 0.29) |
| PS (mL/min/mL) | 0.88 (0.59,1.45) | 3.3 (0.61, 4.74) |

*Table 2: Median (25${}^{th}$ percentile, 75${}^{th}$ percentile) kinetic parameter estimates, on the patient data, using the Bayesian inference and non-linear least squares fitting approaches.*

**Figure 5.**

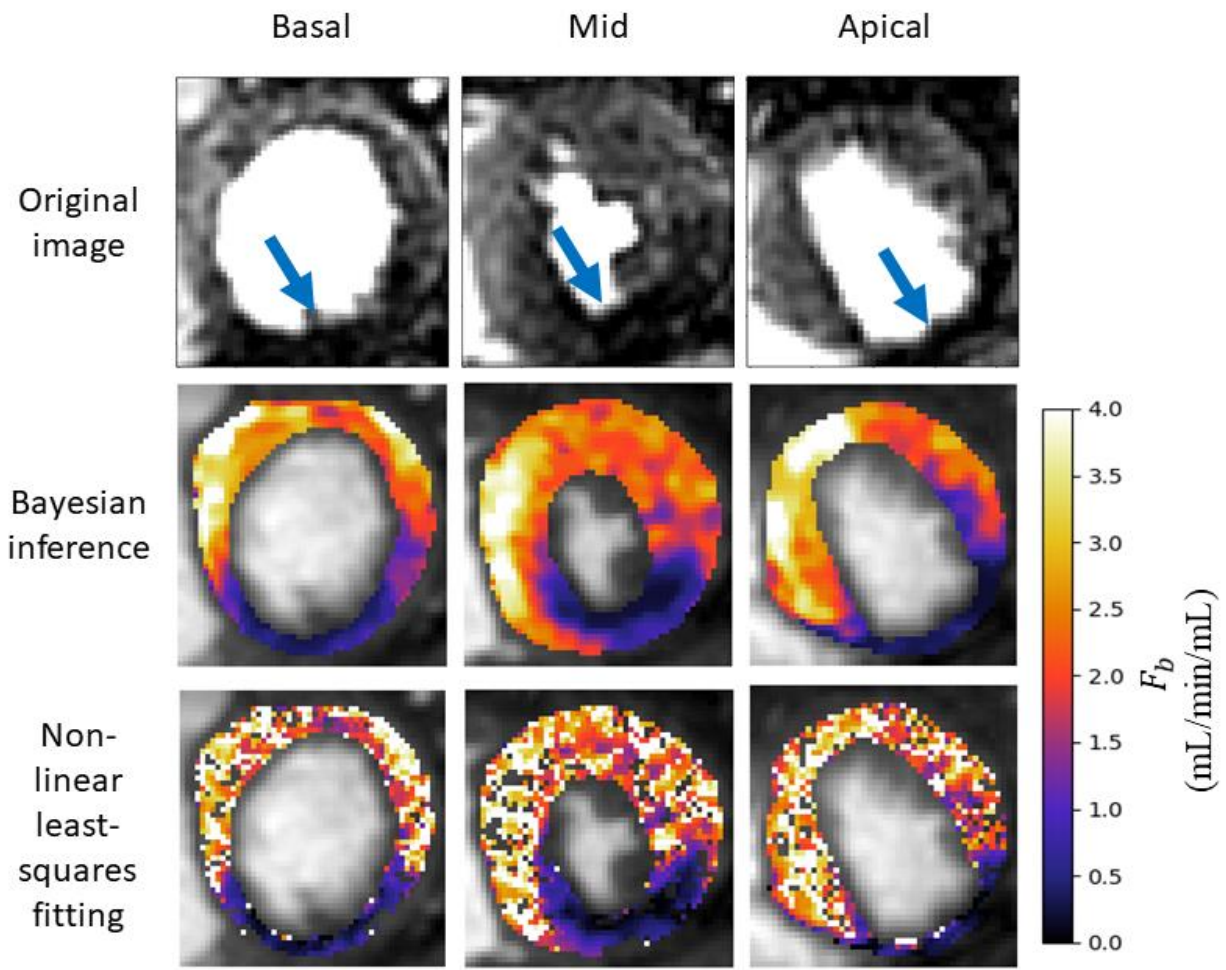

A comparison of Bayesian inference versus least-squares fitting with 100 random initialisations for the three acquired slices for a patient with an overt perfusion defect (as indicated by the arrows in the first row). While both techniques identify the area of ischaemia the least-squares fittings have severe speckle-like noise and even gaps where the fitting has failed. This makes it more difficult to accurately delineate the boundaries of the ischaemic area and can lead to areas where the ischaemia is missed.

**Figure 6.**

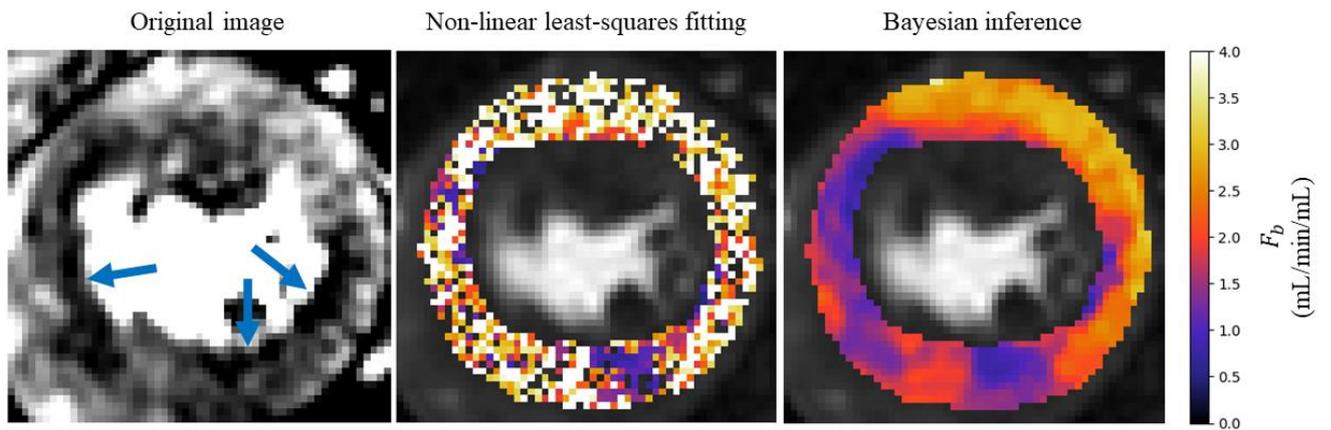

*An example slice where the Bayesian fitting (right) has correctly identified ischaemia but the noisy least squares fitting (middle) makes it difficult to identify the ischaemia, particularly in the infero-septal wall. The identified areas of ischaemia are indicated on the original MR image (left).*

**Figure 7.**

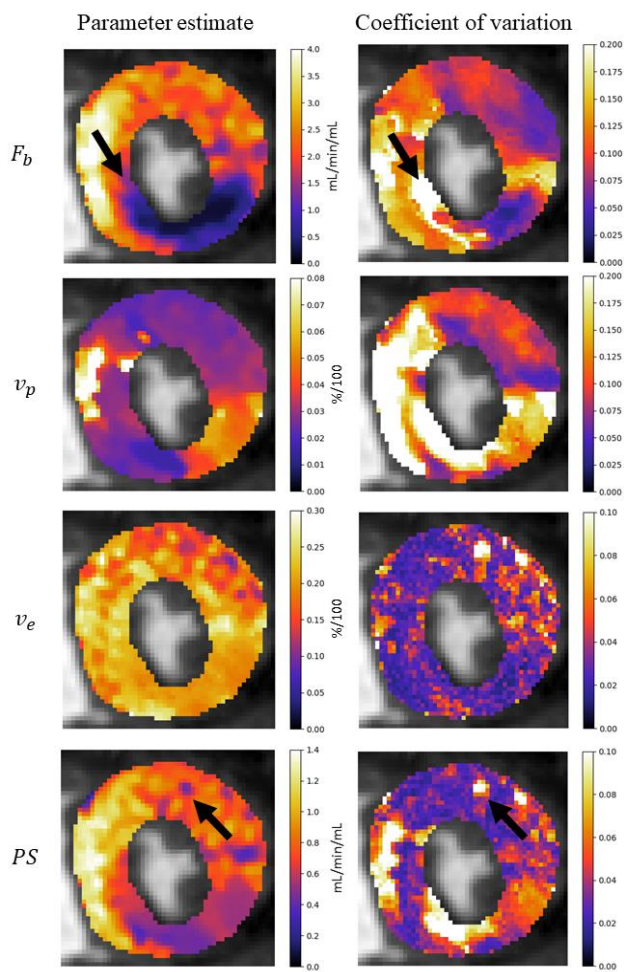

*Parameter maps for the four kinetic parameters of the 2CXM. The coefficient of variation represents the uncertainty about the parameter estimate and could be incorporated into the clinical decision making process. Black arrows are used to compare areas of high uncertainty to the respective parameter estimates.*

## 5. Discussion

In this work, the use of Bayesian inference to estimate tracer-kinetic parameters from myocardial perfusion MRI data is investigated. This approach incorporates both spatial prior knowledge and prior knowledge on the kinetic parameter values. It also enables the computation of posterior distributions over the model parameters. It is compared to the more traditional method of parameter estimation, the non-linear least squares fitting. This comparison first assesses the accuracy and reproducibility of the parameter estimations in a simulated, but realistic, setting. The two methods are also compared using patient data to assess the success of disease detection using the quantitative flow maps.

As discussed, a possible alternative approach is to use a simpler model for the quantitative modelling such as using a Fermi-constrained deconvolution (Jerosch-Herold et al., 1998; Zarinabad et al., 2012). The model to be fit to the data is simpler and has fewer parameters and thus can be fit more reliably, with less frequent failed fittings. However, such an approach only allows the resolution of MBF and the other parameters have no physiological interpretation. It is hypothesised that the extra physiological parameters that can be resolved using the two-compartment exchange model may allow a more informative assessment of the tissue. Tracer-kinetic models can also be fit more reliably on a segment-wise level due to the reduced noise after the signal averaging. However, it has been shown that the reduction of spatial resolution leads to a loss of diagnostic information (Villa et al., 2016; Zarinabad et al., 2015). It is also likely that increasing the number of time points that are sampled may increase the reliability of the estimates but this may not be possible in a clinical setting.

Lehnert et al. have also recently proposed the use of spatial regularisation (Lehnert et al., 2018). In this work, a Tikhonov (L2-norm) regularisation term is added to the cost function to be used in a gradient-based optimisation process. However, this is known, and seen in this work, to introduce smoothing over physiological borders where a large difference in kinetic parameters occurs. In fact, in Kelm et al. (Kelm et al., 2009) it was shown that a L1-norm regularisation is more suitable in applications that possess sharp edges between kinetic parameters, such as myocardial perfusion MRI. This motivates the use of a Laplace prior in our work which is equivalent to L1-norm regularisation. The benefits of the Bayesian approach also include the use of the MCMC exploration of parameter space which is less susceptible to local optima than gradient-based optimisations. Bayesian inference also yields an approximation of the posterior distribution of the parameters rather than a point-estimate with no indication of uncertainty.

Bayesian inference of tracer-kinetic parameters using DCE-MRI has been proposed previously (Dikaios et al., 2017; Orton et al., 2007; Schmid et al., 2006) and has in general been shown to be more reliable than non-linear least squares fitting. This work is however the first application to myocardial perfusion data, to our knowledge. The main innovation of this work is the utilisation of hierarchical priors. As discussed, hierarchical models allow model parameters to vary by group. The effect of using fixed priors is shown in Fig. 4. where the parameter estimates cannot adapt to areas that are largely different from the prior information (for example a perfusion defect). In this application, this is desirable in order to avoid the averaging effects between areas of ischaemia and healthy myocardium without having to distinguish between the two groups *a priori*. Hierarchical modelling has been applied to DCE-MRI data by Schmid et al. (Schmid et al., 2009) in the setting of a clinical trial where two scans were acquired per patient, before and after treatment, leading to a temporal change in the kinetic parameters and thus two distinct groups of patients. This is different to our work which instead treats individual voxels in the spatial context as being from distinct groups, healthy or diseased.

Using the simulations, it is found in this work that the Bayesian inference technique presented is significantly more accurate than the standard least-squares fitting as evidenced by the NMSE between the estimated and true parameter values. Furthermore, the variability of the estimates is reduced, as shown by the lower standard deviation and the estimates are more reproducible across different noise realisations. The benefit of using a hierarchical model is also demonstrated. It is seen that in the presence of areas of reduced MBF, the prior knowledge of stress MBF values does not apply and the non-hierarchical model cannot account for the differences in groups of voxels (ischaemic and healthy). This leads to an averaging of the information from the data and the prior information and thus an over-estimation of MBF in these areas.

Naturally, there are no ground-truth values for comparison with the patient data estimates. However, the Bayesian parameter estimation leads to reduced numbers of outliers and failed fittings as compared to the least-squares fitting. The effect of this is shown in the parameter maps in Fig. 5. In this example, the perfusion defect in the inferior segment of the myocardium is clearly identified using both the Bayesian inference and least-squares fitting. However, there is still some speckle like noise present in the least-squares estimates, even after 100 repeated fittings. The noisy estimates can make it difficult to delineate the boundaries of the ischaemia and in this case lead to the underestimation of the extent of the ischaemia. It is clear from that the Bayesian inference is identifying correctly the area of reduced contrast uptake in the inferior segment, which is visible in the original MR image at the correct windowing level and is easily picked up when assessing the quantitative flow maps.

These findings are in line with previous literature, Broadbent et al. (Broadbent et al., 2013) reported failed fittings in 10% of cases on a segment-wise level. This is despite the fact they considered curves which have been averaged over a segment of the myocardium to boost SNR. Schwab et al. (Schwab et al., 2015) reports a median flow value ($25^{th}$ percentile, $75^{th}$ percentile) of 3.055 (1.197,1168.4) mL/min/mL using the 2CXM model with the conventional least-squares fitting approach. The $75^{th}$ percentile value reported is well in excess of 100 times of the range of values that are physiologically feasible. Both the mean and $75^{th}$ percentile are lower in the results we have presented, due to the bounds used in the optimiser in our implementation but we also found a number of failed fittings and outliers. Furthermore, in this work the *PS* values are extremely variable which could be due to the short acquisition period. Capillary permeability is known to affect the later part of the curves and this process may not be fully observed. This indicates the unreliability of conventional perfusion estimates and hence the difficulty of the clinical translation of quantitative perfusion analysis is apparent.

Conversely, recent work, as presented by Kellman et al. (Kellman et al., 2017) has shown reproducible MBF values, using similar tracer-kinetic models, in a consistent population of healthy volunteers (Brown et al., 2018) and a good correlation with the MBF values derived from positron emission tomography (PET) (Engblom et al., 2017). However, this work did not present a new methodology for the model fitting and thus it is unclear how the fitting issues discussed here were addressed. Furthermore, this work has only considered the MBF and the other kinetic parameters are not validated. As our results in Fig. 2. show, due to parameter coupling it is impossible to judge the reliability of fitting by evaluating only a single parameter. We have demonstrated the ability to robustly infer the kinetic parameters and shown smoothness among all estimated parameters.

The difficulty associated with the least-squares fitting is due to the complex nature of the cost function which can contain many local optima. The gradient-based optimisation schemes are susceptible to converging to the local minima and thus returning inaccurate parameter estimates. This problem is exacerbated by the relative complexity of the 2CXM relative to the observed data and the complex errors introduced by the imaging process. The result of this is the noisy and often inaccurate estimates seen in this study.

Further well known issues with the standard least-squares fitting technique are that it is difficult to assess the uncertainty of the estimates and that these estimates are strongly dependent on the initial conditions of the optimisation process. The latter of these issues can be mitigated by using many randomly chosen initial positions but there is no structured or robust approach to doing this. These issues combine to limit the applicability of quantitative perfusion analysis in a clinical setting. Indeed, the patient data experiments show that the successful clinical classification of patients is worse with the least-squares fitting while perfect results are achieved with the Bayesian inference, albeit with a small sample size. The Bayesian inference does not depend on the initialisation of the optimisation as a burn-in period is used and these sample values are discarded. Furthermore, it provides a natural framework for quantifying the uncertainty of the estimates through the computation of the a posteriori probability distribution of the parameters.

In this work, using Bayesian inference, a median flow value ($25^{th}$ percentile, $75^{th}$ percentile) of 2.35 (1.9, 2.68) mL/min/mL is computed. The $25^{th}$ and $75^{th}$ percentile values are well within the range of what is physiologically feasible, showing the increased reliability of the parameter estimates obtained using Bayesian inference. Despite the fact that these studies have been conducted with different cohort of patients it still serves to show the significant improvement that is gained by employing a Bayesian inference approach to the parameter estimation.

The coefficient of variation of the posterior distribution is used as a measure of uncertainty in the parameter estimate. The reported values indicate a reasonable level of confidence in the parameter estimates with the median coefficient of variation being 6.6%. However, higher coefficients of variation are also found, indicating high uncertainty in some regions. In Fig. 7., in the MBF parameter map ($F_b$), it is seen that there is a high level of uncertainty at the border between the ischaemic and healthy regions (indicated by arrows). It makes sense that there more uncertainty in these border regions and it could possibly be as a result of conflicting information from its neighbouring voxels which could be either ischaemic or healthy. In the $PS$ parameter map, there is also an isolated area of reduced permeability. However, it is seen to be associated with a high level of uncertainty. This uncertainty can be incorporated into an assessment of whether or not there is reduced capillary permeability here. Thus, this uncertainty measure may prove to be useful in the clinical decision-making process but further work on this topic is warranted.

The improved results are due to the MCMC fitting, which better explores parameter space as compared to gradient-based methods, and the use of prior information. Prior information on the kinetic parameters acts as regularisation and constrains the parameter values to physiologically realistic ranges. The spatial prior information increases the amount of information used when fitting each voxel, enforces smoothness and reduces the speckle-like noise in the estimates. However, while spatial smoothness is enforced, it is seen in the figures that the parameter maps still preserves sharp change and there is not an over-smoothing of important physiological information.

## 5. Limitations

One of the main criticisms of MCMC algorithms is the large computational cost involved in accurately approximating the posterior distribution, though the use of multiple random initialisations in the least-squares fitting is similarly computationally expensive. The computational cost could potentially be addressed using an efficient GPU-based implementation.

A limitation of tracer-kinetic modelling, in general, is that the models used are simplified versions of the underlying processes. The aim of this work was to examine whether Bayesian inference can yield more reliable parameter estimates that non-linear least squares fitting with the 2CXM. There has been no effort made to investigate whether this is the most suitable model in this application

There are no ground-truth parameter values for patient data and as such there is no way to comment directly on the accuracy of the parameter estimates. The absolute quantitative accuracy needs validation in comparison to a gold standard technique such as microspheres. Further work is also required on the clinical utility of the findings. In this work, diagnostic accuracy is only compared with the expert clinical assessment, however future work will involve comparisons with the gold standard examinations, invasive coronary angiography and fractional flow reserve, in a larger patient cohort.

## 6. Conclusion

Tracer-kinetic parameters can be accurately and robustly inferred from myocardial perfusion MRI using hierarchical Bayesian inference. The use of a MCMC fitting scheme and the inclusion of spatial prior knowledge improves the reliability of the parameter estimation as compared with least-squares fitting. As a result of the improved model fitting, the diagnostic capabilities of the technique is increased.

## Acknowledgements

The authors acknowledge financial support from the King's College London & Imperial College London EPSRC Centre for Doctoral Training in Medical Imaging (EP/L015226/1); Philips Healthcare; The Department of Health via the National Institute for Health Research (NIHR) comprehensive Biomedical Research Centre award to Guy's & St Thomas' NHS Foundation Trust in partnership with King's College London and King's College Hospital NHS Foundation Trust and via the NIHR Cardiovascular MedTech Co-operative at Guy's and St Thomas' NHS Foundation Trust; The Centre of Excellence in Medical Engineering funded by the Wellcome Trust and EPSRC under grant number WT 088641/Z/09/Z.

# Appendix

*A.1 Residue function*

The pair of coupled differential equations (1), (2) can be solved analytically using the Laplace transform to yield a solution in the form:

$$C_\Theta(t) = R_F(t, \Theta) * C_{AIF}(t - \tau_0)$$

The residue function $R_F$ is given as: $R_F(t, \Theta) = A \exp(\alpha t) + (1 - A) \exp(\beta t)$, where:

$$\alpha, \beta = \frac{1}{2}\left[-\left(\frac{PS}{v_p} + \frac{PS}{v_e} + \frac{F_p}{v_p}\right) \pm \sqrt{\left(\frac{PS}{v_p} + \frac{PS}{v_e} + \frac{F_p}{v_p}\right)^2 - 4\frac{PS}{v_e}\frac{F_p}{v_p}}\right]$$

$$A = \frac{\alpha + \frac{PS}{v_p} + \frac{PS}{v_e}}{\alpha - \beta}$$